\documentclass[twocolumn]{book}
\usepackage{graphicx,color}
\usepackage{makeidx,universe}

\usepackage{amssymb}

\makeauthorindex
\BookTitle{Proceedings of the XXIX PHYSICS IN COLLISION}

\CopyRight{\copyright 2009 by Universal Academy Press, Inc.}

\begin{document} %******************************************

%\tableofcontents
\pagenumbering{arabic}

\chapter{%
{\LARGE \sf
The $f_{D_s}$ Puzzle} \\
{\normalsize \bf %%%%%%%%%%%%%%******* Authors **************
Andreas S. Kronfeld} \\
{\small \it \vspace{-.5\baselineskip}%***** Affiliations ***********
Theoretical Physics Department, Fermi National Accelerator Laboratory,
    P.O. Box 500, Batavia, IL 60510, USA}
}

%**************************
% Please note:
% One \AuthorContents{} is necessary
% for EACH CONTRIBUTION, for the contents page and
% One \AuthorIndex{} is necessary
% for EACH AUTHOR, for the index.
%**************************

%***** Item below is the data for CONTENTS. 
%***** Please enter all author's name that should be initialized.
\AuthorContents{A.S. Kronfeld}

%***** items below are the data for AUTHOR INDEX. 
%***** Please enter a author's name that should be initialized.
\AuthorIndex{Kronfeld}{A.S.} 

\baselineskip=10pt %*******
\parindent=10pt    %*******

\section*{Abstract}

Recent measurements of the branching fraction for $D_s\to\ell\nu$ 
disagree with the Standard Model by around $2\sigma$.
In this case the key aspect of the Standard Model is the calculation of 
the decay constant, $f_{D_s}$, with lattice QCD.
This talk surveys the experimental measurements, and explains how the 
lattice QCD calculations are done.
Should the discrepancy strengthen again (it was earlier $3.8\sigma$), 
it would be a signal of new physics.
Models that could explain such an effect are also discussed.       

\section{Introduction}
\label{sec:intro}

The decay constant of a meson parametrizes the decay of the meson to 
leptons.
For a pseudoscalar like the charmed strange meson~$D_s$, it is 
defined by the hadronic matrix elements
\begin{eqnarray}
    \langle 0|\bar{s}\gamma_\mu\gamma_5c|D_s\rangle & 
        \hspace*{-6pt} = \hspace*{-6pt} & if_{D_s}p_\mu, 
    \label{eq:kronfeld:AfDs} \\
    (m_c+m_s)\langle 0|\bar{s}\gamma_5c|D_s\rangle & 
        \hspace*{-6pt} = \hspace*{-6pt} & -m_{D_s}^2f_{D_s}.
    \label{eq:kronfeld:PfDs}
\end{eqnarray}
The Ward identity of the partially conserved axial current (PCAC) 
ensures that the two definitions are identical.
These matrix elements are directly computable in QCD, via numerical 
simulations of lattice gauge theory.
These calculations are useful: the ratio $f_\pi/f_K$, for example, is 
used to determine the Cabibbo angle, $\tan\theta_C\propto%
[B(K\to l\nu)/B(\pi\to l\nu)]^{1/2}f_\pi/f_K$~\cite{Marciano:2004uf}.

In the Standard Model, the expression for the branching ratio is
\begin{equation}
    B(D_s\to\ell\nu) = \frac{m_{D_s}\tau_{D_s}}{8\pi} f_{D_s}^2
        \left(1-\frac{m_\ell^2}{m_{D_s}^2}\right)^2
        \left|G_FV_{cs}^*m_\ell\right|^2,
    \label{eq:kronfeld:BrSM}
\end{equation}
with analogous expressions for leptonic decays of other pseudoscalar 
mesons.
Here $G_F$ is the Fermi constant, measured in muon decay, and $V$ is 
the CKM matrix.
Beyond the Standard Model, one replaces
\begin{equation}
    G_FV_{cs}^*m_\ell \mapsto G_FV_{cs}^*m_\ell +
        G_A^\ell m_\ell + G_P^\ell\frac{m_{D_s}^2}{m_c+m_s},
    \label{eq:kronfeld:ampBSM}
\end{equation}
where $G_A^\ell$ and $G_P^\ell$ are related to couplings and masses 
of new interactions in a way analogous to $G_F=g^2/\sqrt{2}M_W^2$.
Amplitudes that proceed through an axial-vector current ($G_F$ and 
$G_A$) are helicity suppressed, but amplitudes that proceed through 
a pseudoscalar current ($G_P$) are not.
In practice, $D_s\to\mu\nu$ has a helicity-suppression factor 
$m_\mu^2/m_{D_s}^2=2.8\times10^{-3}$.
On the other hand, $D_s\to\tau\nu$ is not helicity suppressed but, 
instead, phase-space suppressed:
$(1-m_\tau^2/m_{D_s}^2)^2=3.4\times10^{-2}$.

The decay constant of the $D_s$ was expected to be an excellent test of 
lattice QCD, for several reasons~\cite{Briere:2001rn}.
The matrix elements in Eqs.~(\ref{eq:kronfeld:AfDs}) 
and~(\ref{eq:kronfeld:PfDs}) are gold-plated, in the sense of 
Ref.~\cite{Davies:2003ik}, namely, only one hadron enters, and the 
chiral extrapolation is controlled.
Experimental measurements of $|V_{cs}|f_{D_s}$, via 
Eq.~(\ref{eq:kronfeld:BrSM}), can be combined with the determination 
of $|V_{cs}|$ from CKM unitarity.
The idea that new physics could compete is usually discounted, 
because the decay is Cabibbo-favored and proceeds at the tree level 
of the weak interactions.
Finally, the precision of experiments has lagged that of calculations, 
so the analysis of the numerical lattice-QCD data is carried out with a 
relatively blind eye.

The first round of testing seemed to go well.
In June 2005, the first lattice-QCD calculation with 2+1 flavors of 
sea quarks appeared in a joint work of the Fermilab Lattice and MILC 
Collaborations:
\begin{equation}
    f_{D_s} = 249\pm 3 \pm 16~\textrm{MeV}~\textrm{Fermilab/MILC}~%
        \cite{Aubin:2005ar},
    \label{eq:kronfeld:prl05}
\end{equation}
where the first error is statistical and the second systematic.
This prediction was followed a year later by a comparably accurate
measurement of $B(D_s\to\mu\nu)/B(D_s\to\phi\pi)$ that, when 
combined with an independent measurement of $B(D_s\to\phi\pi)$, 
yielded
\begin{eqnarray}
    \hspace*{-1.0em}
    f_{D_s} = 283 \pm 17 \pm 7 \pm 14~\textrm{MeV} & 
        \hspace*{-0.5em} \mu\nu/\phi\pi \hspace*{-0.5em} &
        \textrm{BaBar}~\cite{Aubert:2006sd},
    \label{eq:kronfeld:prl06}
\end{eqnarray}
which agrees with Eq.~(\ref{eq:kronfeld:prl05}) at 1.2$\sigma$.
Both results are discussed in more detail below.

But then something unexpected happened.
During 2007 the CLEO and Belle Collaborations both published absolute 
measurements of $B(D_s\to\mu\nu)$~\cite{Pedlar:2007za,Widhalm:2007mi}, 
and CLEO also published absolute measurements of 
$B(D_s\to\tau\nu)$~~\cite{Pedlar:2007za,Ecklund:2007zm}.
Transcribed via Eq.~(\ref{eq:kronfeld:BrSM}) as measurements of the 
decay constant, these are
\begin{eqnarray}
    f_{D_s} = 264\pm15\pm\;7~\textrm{MeV} & \mu\nu  & 
        \textrm{CLEO}~\cite{Pedlar:2007za}, 
    \label{eq:kronfeld:cleo07munu} \\
    f_{D_s} = 275\pm16\pm 12~\textrm{MeV} & \mu\nu  & 
        \textrm{Belle}~\cite{Widhalm:2007mi}, \\
    f_{D_s} = 310\pm25\pm\;8~\textrm{MeV} & \tau\nu & 
        \textrm{CLEO}~\cite{Pedlar:2007za}, 
    \label{eq:kronfeld:cleo07pinunu} \\
    f_{D_s} = 273\pm16\pm\;8~\textrm{MeV} & \tau\nu & 
        \textrm{CLEO}~\cite{Ecklund:2007zm}.
    \label{eq:kronfeld:cleo07enununu}
\end{eqnarray}
Taking Eqs.~(\ref{eq:kronfeld:prl06})--(\ref{eq:kronfeld:cleo07enununu}) 
at face value, the weighted average (combining all errors in quadrature)
is~\cite{Dobrescu:2008er}
\begin{equation}
    f_{D_s} = 277 \pm 9~\textrm{MeV},
    \label{eq:kronfeld:avg200803}
\end{equation}
which is $1.5\sigma$ higher than the value in 
Eq.~(\ref{eq:kronfeld:prl05}).
Meanwhile, and more dramatically, the HPQCD Collaboration published 
a lattice-QCD calculation with an error significantly smaller than 
Fermilab/MILC's:
\begin{equation}
    f_{D_s} = 241 \pm 3~\textrm{MeV}~\textrm{HPQCD}~%
        \cite{Follana:2007uv}.
    \label{eq:kronfeld:hpqcd}
\end{equation}
Section~\ref{sec:kronfeld:lattice} explains why the error is so much 
smaller.
The difference between Eqs.~(\ref{eq:kronfeld:avg200803}) 
and~(\ref{eq:kronfeld:hpqcd}) is $3.8\sigma$.
(Omitting BaBar, as in Ref.~\cite{Amsler:2008zzb}, the discrepancy 
becomes $3.4\sigma$.)
It is important to bear in mind---and easy to see from
Eqs.~(\ref{eq:kronfeld:avg200803}) and~(\ref{eq:kronfeld:hpqcd})---that 
the yardstick for $\sigma$ is the experimental statistical error.
The 2008 edition of the Review of Particle 
Physics~\cite{Amsler:2008zzb} noted that the discrepancy could be a 
sign of physics beyond the Standard Model.
Candidate models are discussed below.

The rest of this paper brings this saga up to date.
Section~\ref{sec:kronfeld:expt}\ gives a brief survey of the 
experiments, including higher-statistics measurements from CLEO, and 
a (different) extraction of $f_{D_s}$ from BaBar's measurement of
$B(D_s\to\mu\nu)/B(D_s\to\phi\pi)$ by the Heavy Flavor Averaging 
Group (HFAG), which have brought the discrepancy down to $2.3\sigma$.
Section~\ref{sec:kronfeld:lattice}\ discusses recent developments in 
lattice-QCD calculations.
New physics explanations are in Sec.~\ref{sec:kronfeld:new}~~
The main issues are summarized in Sec.~\ref{sec:kronfeld:summary}

\section{Measurements}
\label{sec:kronfeld:expt}

Observations of $D_s\to\ell\nu$ date back to 1993 in fixed-target 
experiments and $e^+e^-$ collisions~\cite{Aoki:1992wj,Acosta:1993rx,%
Bai:1994qz,Kodama:1996xq,Acciarri:1996bv,Chadha:1997zh,%
Alexandrov:2000ns,Abbiendi:2001nb,Heister:2002fp}.
These early measurements are omitted from (the current) Particle Data 
Group (PDG) and HFAG averages and are also omitted from this discussion.
In 2006, the PDG~\cite{Yao:2006px} included a correlated average of 
Refs.~\cite{Acciarri:1996bv,Chadha:1997zh,Alexandrov:2000ns,%
Abbiendi:2001nb,Heister:2002fp}, which increase the discrepancy, 
as discussed below, by 0.3--0.4$\sigma$.

\subsection{CLEO $\mu\nu$ and $\tau\nu$}
\label{sec:kronfeld:cleo-mu}

CLEO produces $D_sD_s^{(*)}$ pairs in $e^+e^-$ collisions just above 
threshold, as in a 1994 observation by the BES 
Collaboration~\cite{Bai:1994qz}.
The multiplicity is low, so the whole event can be reconstructed, and 
the neutrino is ``detected'' by requiring the missing mass-squared to 
be consistent with~0.
Radiative events with photon energy greater than 300~MeV are rejected.
Although this cut is imposed for other reasons, it usefully
removes radiative events without helicity suppression.

In $D_s\to\tau\nu$, the $\tau$ decays in the detector, and the details 
of the analyses depend on the $\tau$-decay mode.
CLEO first observed events in which $\tau\to\pi\nu$ as a background to 
the $D_s\to\mu\nu$ analysis, but then turned these events into a 
measurement.
A separate analysis chain counts $D_s\to\tau\nu$ events in which 
$\tau\to e\nu\nu$.
With 2 or 3 neutrinos in the final state, the constraint on missing 
mass-squared is no longer pertinent.
These analyses also reject events with photons, but this is a matter 
of $\tau$ detection.
In the $D_s$ rest frame, the $\tau$ acquires only 9.3~MeV of kinetic 
energy, so radiative events are not an issue.

In January 2009, CLEO published analyses with their full 
data-set, reporting
\begin{eqnarray}
    \hspace*{-2.0em} 
    f_{D_s} = 257.3\pm 10.3\pm 3.9~\textrm{MeV} & \hspace*{-0.5em}
        \mu\nu & \hspace*{-0.5em} \cite{Alexander:2009ux}, \\
    \hspace*{-2.0em} 
    f_{D_s} = 278.7\pm 17.1\pm 3.8~\textrm{MeV} & \hspace*{-0.5em}
        \tau\nu,~\tau\to\pi\nu & \hspace*{-0.5em} \cite{Alexander:2009ux},
    \label{eq:kronfeld:cleo09pinunu} \\
    \hspace*{-2.0em} 
    f_{D_s} = 252.5\pm 11.1\pm 5.2~\textrm{MeV} & \hspace*{-0.5em} 
        \tau\nu,~\tau\to e\nu\nu & \hspace*{-0.5em} \cite{Onyisi:2009th},
    \label{eq:kronfeld:cleo09enununu}
\end{eqnarray}
which supersede Eq.~(\ref{eq:kronfeld:cleo07munu}),
(\ref{eq:kronfeld:cleo07pinunu}) 
and~(\ref{eq:kronfeld:cleo07enununu}), respectively.
After Physics in Collision 2009, CLEO made public an analysis of a 
third $D_s\to\tau\nu$ decay chain, $\tau\to\rho\nu$, yielding
\begin{eqnarray}
    \hspace*{-2.0em} 
    f_{D_s} = 257.8\pm 13.3\pm 5.2~\textrm{MeV} & \hspace*{-0.5em}
        \tau\nu,~\tau\to\rho\nu & \hspace*{-0.5em} \cite{Naik:2009tk}.
    \label{eq:kronfeld:cleo09rhonunu}
\end{eqnarray}
A novelty of this analysis is that it disentangles a mesa-shaped 
signal distribution from a peaking background.

\subsection{BaBar and Belle $\mu\nu$}
\label{sec:kronfeld:Bfactories}

BaBar and Belle, following a strategy devised by 
CLEO~\cite{Acosta:1993rx,Chadha:1997zh}, collect a $D_s$ sample from 
continuum events under the $\Upsilon(4S)$ by observing the decay 
$D_s^*\to D_s\gamma$.
BaBar then counts the relative number of $D_s\gamma\to\mu\nu\gamma$ and 
$D_s\gamma\to\phi\pi\gamma$ events, yielding a measurement of 
$B(D_s\to\mu\nu)/B(D_s\to\phi\pi)$.
A~separate measurement of $B(D_s\to\phi\pi)$ is needed to extract 
$f_{D_s}$ via Eq.~(\ref{eq:kronfeld:BrSM}), and BaBar used an average 
of two of its own measurements~\cite{Aubert:2005xu}.
Belle improves on the $D_s^*\to D_s\gamma$ technique by devising a 
Monte Carlo analysis to guide full reconstruction of the event.
In this way they obtain an absolute measurement of $B(D_s\to\mu\nu)$.

Measurements of $B(D_s\to\mu\nu)/B(D_s\to\phi\pi)$ are subject to some 
ambiguity.
The $\phi$ decays to $KK$, but other processes, such as 
$D_s\to f_0\pi\to KK\pi$ also contribute.
The two contributions are not completely separable, because the 
amplitudes interfere~\cite{Alexander:2008cq}.

\subsection{CKM; Radiative Corrections}
\label{sec:kronfeld:ckm}

To extract $f_{D_s}$ from the measurements of the branching ratio, 
one needs a value of the CKM matrix element~$|V_{cd}|$.
In practice, it has been determined from CKM unitarity, either using a 
global fit or simply setting $|V_{cs}|=|V_{ud}|$.
(It makes an insignificant difference.)
With four or more generations, this assumption incorrect, but
4- (or more) generation CKM unitarity still requires $|V_{cs}|\le1$.
Therefore, an incorrect assumption about $|V_{cs}|$ cannot explain 
why the ``measured'' value of $f_{D_s}$ is too high.

Leptonic decays are, of course, subject to radiative corrections.
A~class of virtual processes are of special interest here, namely
$D_s\to D_s^*\gamma\to\mu\nu\gamma$, where $D_s^*$ is a vector or 
axial-vector meson.
The decay $D_s^*\to\mu\nu$ is not subject to helicity suppression, so 
the absence in the rate of a factor $(m_\mu/m_{D_s})^2$ could 
compensate for the presence of the factor $\alpha\approx1/137$.
The radiative rate is significant for energetic 
photons~\cite{Burdman:1994ip,Hwang:2005uk}.
With CLEO's cut rejecting radiative events with $E_\gamma>300$~MeV, 
however, Eq.~(12) of Ref.~\cite{Burdman:1994ip} shows that these events 
add only around 1\% to the rate and, thus, cannot be an explanation of 
the discrepancy.

\subsection{HFAG}
\label{sec:kronfeld:hfag}

The experimental collaborations' differing treatments of $|V_{cs}|$ and 
of radiative corrections are not yet signifigant, so 
Refs.~\cite{Dobrescu:2008er,Amsler:2008zzb} simply average quoted 
values of $f_{D_s}$.
Eventually, however, a uniform treatment will be necessary, so,
with this in mind, the Heavy Flavor Averaging Group (HFAG) 
\cite{Schwartz:2009hv,HFAG} has undertaken to average the 
model-independent quantities $B(D_s\to\mu\nu)$, 
$B(D_s\to\tau\nu)$, and $B{(D_s\to\mu\nu)}/B(D_s\to\phi\pi)$.
The averaging is straightforward.
When turning to the extraction of $f_{D_s}$, however, HFAG noticed an 
important issue with BaBar's determination of $f_{D_s}$. 
The definition of the $\phi$ resonance in Ref.~\cite{Aubert:2006sd} is 
a window of $K^+K^-$ invariant mass $M_{K^+K^-}$, such that 
$|M_{K^+K^-}-m_\phi|<5.5~\textrm{MeV}$.
The normalizing measurements, on the other hand, 
used ${|M_{K^+K^-}-m_\phi|}<15~\textrm{MeV}$ \cite{Aubert:2005xu}.
From the $M_{K^+K^-}$ distribution in Ref.~\cite{Alexander:2008cq}, 
it is clear that the difference is important.
Fortunately, CLEO~\cite{Alexander:2008cq} reports 
$B(D_s\to K^+K^-\pi)$ as a function of $M_{K^+K^-}$, so HFAG combines 
$B(D_s\to K^+K^-\pi)$ with $|M_{K^+K^-}-m_\phi|<5~\textrm{MeV}$,
$B(\phi\to K^+K^-)$, and BaBar's $B(D_s\to\mu\nu)/B(D_s\to\phi\pi)$ 
to arrive at $B(D_s\to\mu\nu)$.
Interpreting this branching fraction as $f_{D_s}$ yields
\begin{eqnarray}
    \hspace*{-1.5em}
    f_{D_s} = 237.3 \pm 16.7 \pm 1.7~\textrm{MeV} & 
        \hspace*{-0.5em} \mu\nu/\phi\pi \hspace*{-0.5em} &
        \textrm{HFAG}~\cite{HFAG},~~
    \label{eq:kronfeld:HFAG}
\end{eqnarray}
which we shall use to supersede Eq.~(\ref{eq:kronfeld:prl06}).
It is 16\% or 2.9$\sigma$ lower (using the normalization and 
systematics for~$\sigma$).

\subsection{Synopsis}
\label{sec:kronfeld:synopsis}

In summary, the measurements of the branching fraction 
$B(D_s\to\ell\nu)$ are relatively straightforward counting experiments.
They can be contrasted with, say, searches for the Higgs boson at 
hadron colliders~\cite{Peters:2009jy}, in which a careful and subtle 
modeling of the QCD background is essential.
Here the background is small and-or measurable;
the events are clean, or even pristine.
As a result, the dominant experimental error is statistical.
It is, of course, possible that more experiments have fluctuated
up than~down.

With the new results, including Eqs.~(\ref{eq:kronfeld:cleo09rhonunu}) 
and~(\ref{eq:kronfeld:HFAG}), the experimental average is now (I find)
\begin{equation}
    f_{D_s} = 257.8 \pm 5.9~\textrm{MeV},
    \label{eq:kronfeld:avg200911}
\end{equation}
or 1.7$\sigma$ lower than Eq.~(\ref{eq:kronfeld:avg200803}),
which is a combination of 1.3$\sigma$ from CLEO's new measurements and 
$1.1\sigma$ from HFAG's revision of BaBar's measurement.
(My average without Eq.~(\ref{eq:kronfeld:cleo09rhonunu}) is 
$257.8\pm 6.4~\textrm{MeV}$, which is close to HFAG's more 
rigorous average of the same inputs, $256.9\pm 6.8~\textrm{MeV}$
\cite{HFAG}.)

\section{Lattice QCD}
\label{sec:kronfeld:lattice}

Lattice QCD has made great strides in the past several 
years~\cite{Davies:2003ik,Bazavov:2009bb}, compared to, say, the 
status at Physics in Collision 2002~\cite{Kronfeld:2002nm}.
The key development has been the inclusion of sea quarks, first with 
$n_f=2$ and, then with $n_f=2+1$.
The latter notation means that one sea quark has a mass nearly equal to 
that of the strange quark, and the other two vary over a range 
$0.1m_s\lesssim m_q\lesssim0.5m_s$, such that chiral perturbation 
theory can be used to reach the up- and down-quark masses.

That said, there have been only two calculations of $f_D$ and $f_{D_s}$ 
with $n_f=2+1$ flavors of sea quarks~\cite{Aubin:2005ar,Follana:2007uv},
one of which dominates the average.
Moreover, both use the same ensembles of lattice gauge 
fields~\cite{Bazavov:2009bb,Bernard:2001av}, which have been generated 
using ``rooted staggered fermions'' for the sea quarks.
The rooting procedure leads to some difficulties~\cite{Kronfeld:2007ek}
that are expected to go away in the continuum limit~\cite{Shamir:2004zc}.

The reason for the rooting is that lattice fermion fields correspond to 
more than one species in the continuum limit~\cite{Nielsen:1980rz}.
With staggered fermions there are four species~\cite{Susskind:1976jm}. 
Sea quarks are represented by a determinant of the (lattice) Dirac 
operator, so to reduce 4 species to~1, one can make the 
\emph{Ansatz}~\cite{Hamber:1983kx}
\begin{equation}
    \left[\det_4({\rm stag}+m)\right]^{1/4} \doteq 
        \det_1(D\kern -0.65em /\kern +0.35em + m),
\end{equation}
where the subscript denotes the number of flavors.
The fourth-root can be built into chiral perturbation 
theory~\cite{Aubin:2003mg}.
In fact, in this context the $\frac{1}{4}$ can be replaced by a free 
parameter, which is then fit.
The fit yields $0.28\pm0.03$~\cite{Bernard:2007ps}, 
in excellent agreement with~$\frac{1}{4}$.

The two principle methodological reasons why the error in 
Eq.~(\ref{eq:kronfeld:hpqcd}) is smaller than in 
Eq.~(\ref{eq:kronfeld:prl05}) is that Ref.~\cite{Follana:2007uv} 
treats the charmed quark as a staggered quark~\cite{Follana:2006rc}, 
using a pseudoscalar density with an absolute normalization via a PCAC 
relation~\cite{Smit:1987zh}, and enabling an extrapolation to the 
continuum limit.
By contrast, Ref.~\cite{Aubin:2005ar} treats the charmed quark as a 
heavy quark~\cite{ElKhadra:1996mp}; the current requires a matching 
factor computed in perturbative QCD~\cite{Harada:2001fi}, and the 
discretization effects are (conservatively) estimated with 
power-counting estimates~\cite{Kronfeld:2003sd}.

To tackle charm on currently available lattices, the HPQCD 
Collaboration has developed a highly-improved staggered quark action 
(HISQ), first used to study charmomium~\cite{Follana:2006rc}.
Some of their other results are tabulated in 
Table~\ref{tbl:kronfeld:hpqcd-test}~~
\begin{table}
    \caption[tbl:kronfeld:hpqcd-test]{Results from 
        Refs.~\cite{Follana:2007uv,Follana:2006rc} other than $f_{D_s}$.
        Experimental quantities are taken from Ref.~\cite{Amsler:2008zzb},
        except for $f_{D^+}$, which is from Ref.~\cite{:2008sq}.
        Here $\Delta_q=2m_{D_q}-m_{\eta_c}$, $q=d,s$.}
    \label{tbl:kronfeld:hpqcd-test}
    \centering
    \begin{tabular}{cccc}
        \hline\hline
        Quantity & Expt & HPQCD & units \\
        \hline
        $m_{J/\psi}-m_{\eta_c}$ & $116.4\pm1.2$ & $111\pm5$  & MeV \\
        $\Delta_d$ & 758.7 & $755\pm14$ & MeV \\
        $\Delta_s$ & 956.5 & $944\pm12$ & MeV \\
        $\Delta_s/\Delta_d$ & $1.261\pm0.002$ & $1.252\pm0.015$ & \\
        $f_\pi$   & $130.7\pm0.4$ & $132\pm2$ & MeV \\
        $f_K$     & $159.8\pm0.5$ & $157\pm2$ & MeV \\
        $f_{D^+}$ & $205.8\pm8.9$ & $207\pm4$ & MeV \\
        \hline\hline
    \end{tabular}
\end{table}
Especially noteworthy here is the value of $f_{D^+}$, which agrees with 
CLEO's later measurement~\cite{:2008sq}.
Most effects that would bring Eq.~(\ref{eq:kronfeld:hpqcd}) into 
better agreement with the measurements of~$f_{D_s}$ would also alter 
$f_{D^+}$, spoiling its agreement.

For the $\pi$, $K$, and $D^+$ decay constants, both the chiral and 
continuum extrapolations are crucial.
For the $D_s$, however, the valence charmed and strange quarks ensure 
a mild chiral extrapolation.
The continuum extrapolation turns out to be interesting:
reading values for $f_{D_s}$ off of plots in 
Ref.~\cite{Follana:2007uv}, I have verified the continuum extrapolation 
and found that the slope in $a^2$ conforms with expectations of 
discretization effects of order $\alpha_sa^2m_c\Lambda$.

The relevant portion of HPQCD's error budget is presented in 
Table~\ref{tbl:kronfeld:hpqcd-budget}~~
\begin{table}
    \caption[tbl:kronfeld:hpqcd-budget]{Error budget from 
        Ref.~\cite{Follana:2007uv}.
        Entries in percent.}
    \label{tbl:kronfeld:hpqcd-budget}
    \centering
    \begin{tabular}{lccc}
        \hline\hline
        Source & $f_{D_s}$ & $f_D$ & $f_{D_s}/f_D$ \\
        \hline
        Statistics & 0.6 & 0.7 & 0.5 \\
        Scale $r_1$ & 1.0 & 1.4 & 0.4 \\
        Continuum limit & 0.5 & 0.6 & 0.4 \\
        Chiral limit & 0.3 & 0.4 & 0.2 \\
        Adjust $m_s$ & 0.3 & 0.3 & 0.3 \\
        Adjust $m_d$ $\oplus$ QED & 0.0 & 0.1 & 0.1 \\
        Finite volume & 0.1 & 0.3 & 0.3 \\
        \hline\hline
    \end{tabular}
\end{table}
Most of the row headings are self-explanatory, except for 
``scale~$r_1$,'' which is discussed below.
\pagebreak
The error budget is nearly complete, in my opinion, more complete than 
many error budgets in the lattice-QCD literature.
It does, however, fail to quote an uncertainty for quenching the 
charmed sea.
This is surely a small effect, of order $\alpha_s(\Lambda/m_c)^2$, but 
perhaps commensurate with the $\frac{1}{2}\%$ errors included in 
Table~\ref{tbl:kronfeld:hpqcd-budget}

I shall now discuss $r_1$ in several steps, first motivating why it is 
used, then giving its definition and its value circa 2007.
Being based on an expanding set of numerical data, its value has now 
changed, so I discuss how it affects charmed-meson decay constants.

Lattice gauge theory has a built-in ultraviolet cutoff---the lattice 
itself.
The natural output is a dimensionless number, with physical dimensions 
balanced by powers of the lattice spacing~$a$.
With a decay constant~$f$, one computes $af$ and then must introduce a 
definition for~$a$.
This is necessary not merely to quote a final result in MeV, but also 
to combine calculations at varying~$a$, 
which is needed to understand the continuum limit.
This is done by picking some fiducial mass $M$, and defining 
$a=(aM)_{\rm lat}/M_{\rm expt}$.
This step eliminates one of the free parameters of 
QCD, namely, the bare coupling.

To keep a long story short, no quantity is ideally suited to play the 
role of~$M$.
A popular choice is $1/r_1$, defined 
via~\cite{Sommer:1993ce,Bernard:2000gd} 
\begin{equation}
    r_1^2F(r_1) = 1,
\end{equation}
where $F(r)$ is the force between two static sources of color,
distance $r$ apart.
The advantages of $r_1$ are that it is easy to compute in lattice QCD, 
and that it depends weakly on sea-quark masses and not at all on 
valence-quark masses.
Then one can combine data from several lattices for $r_1f=(r_1/a)(af)$ 
in the chiral and continuum extrapolations.
Other choices of $M$ could complicate these steps.

Of course, $r_1$ is unknown---it cannot be measured in the lab.
It is inferred from the chiral and continuum limit of other quantities.
Reference~\cite{Follana:2007uv} used the value
\begin{equation}
    r_1 = 0.321\pm 0.005~\textrm{fm},
    \label{eq:kronfeld:hpqcd-r1}
\end{equation}
based on MILC's calculations of $r_1/a$~\cite{Bernard:2001av} and 
HPQCD's own calculations of 
$a(M_{\Upsilon(2S)}-M_{\Upsilon(1S)})$ \cite{Gray:2005ur}. 
The 1.6\% uncertainty in $r_1$ translates into a 1.0\% uncertainty on 
$f_{D_s}$ (c.f.\ Table~\ref{tbl:kronfeld:hpqcd-budget}), 
because when $r_1$ varies, the bare valence quark masses inside the 
$D_s$ do too.

This retuning when $r_1$ changes has been studied by the 
Fermilab Lattice and MILC Collaborations 
(although not all details are as yet public).
Since Ref.~\cite{Aubin:2005ar} was published, MILC has extended the 
ensembles to higher statistics, so Fermilab/MILC's decay constant 
analysis has continued, to reduce the total error.
At Lattice 2008 some of the discretization errors were brought under 
better control, leading to $249\pm 11$~MeV \cite{Bernard:2009wr}, with 
(serendipitously) the same central value as 
Eq.~(\ref{eq:kronfeld:prl05}).
References~\cite{Aubin:2005ar,Bernard:2009wr} used 
$r_1=0.318\pm 0.007~\textrm{fm}$~\cite{Bazavov:2009fDs} based on 
essentially the same input information as 
Eq.~(\ref{eq:kronfeld:hpqcd-r1}).

Meanwhile, however, evidence has begun to accumulate that $r_1$ should 
be smaller.
Focusing on MILC's latest analysis of $r_1f_\pi$~\cite{Bazavov:2009fk}, 
one has
\begin{equation}
    r_1 = 0.3108\pm 0.0022~\textrm{fm}.
    \label{eq:kronfeld:fnal-r1}
\end{equation}
Retuning the quark masses, this changes $f_{D_s}$ to
(preliminary, presented at Lattice 2009)
\begin{equation}
    f_{D_s} = 260\pm 10~\textrm{MeV}~\textrm{Fermilab/MILC}~%
        \cite{Bazavov:2009fDs},
    \label{eq:kronfeld:lat09}
\end{equation}
in which 4.2~MeV of the increase stems from the change in $r_1$, and 
the rest from other refinements of the analysis~\cite{Bazavov:2009fDs}.
In other words, a shift down of 2.3\% in $r_1$ has led to a shift up of 
1.7\% in $f_{D_s}$.

The HPQCD Collaboration has also incorporated the extensions of the 
MILC ensembles into its analysis of $r_1$~\cite{Davies:2009ts}.
They find
\begin{equation}
    r_1 = 0.3133\pm 0.0023~\textrm{fm},
\end{equation}
which is 2.4\% lower than the value in Eq.~(\ref{eq:kronfeld:hpqcd-r1}).
Although this suggests an increase in $f_{D_s}$ of 3--5~MeV, one should 
keep in mind that HPQCD's calculations of $f_{D_s}$ have proceeded to 
yet finer lattices.
It seems prudent to wait for their own update, rather than applying a 
shift.

Because both Fermilab/MILC and HPQCD use the same ensembles of lattice 
gauge fields, it is, or should be, a high priority to compute $f_D$ and 
$f_{D_s}$ with other formulations of sea quarks.
A promising development comes from the European Twisted-Mass 
Collaboration (ETMC), which has ensembles with $n_f=2$ over a range of 
sea-quark masses and lattice spacings 
(although not as extensive as MILC's with $n_f=2+1$).
They find $f_{D_s}=244\pm 8$~MeV \cite{Blossier:2009bx}, where the 
error stems from a thorough analysis of all uncertainties \emph{except} 
the quenching of the strange quark.
It is not easy to estimate this error reliably enough for averaging.
Earlier results with $n_f=2$ obtained similar central 
values~\cite{Bernard:2002pc}, or a bit higher~\cite{AliKhan:2000eg}, 
albeit with larger error bars.

It seems reasonable, then, to take as the current best estimate from 
lattice QCD, the weighted average of Eqs.~(\ref{eq:kronfeld:hpqcd}) 
and~(\ref{eq:kronfeld:lat09}):
\begin{equation}
    f_{D_s} = 242.6\pm 2.9~\textrm{MeV}\quad\textrm{LQCD 2009},
\end{equation}
which is 2.3$\sigma$ lower than the average of measurements in
Eq.~(\ref{eq:kronfeld:avg200911}).
The experimental statistical error continues to dominate this~$\sigma$,
although if the central value of the lattice average were to 
increase by 3--5~MeV, the discrepancy would soften below 2$\sigma$.

\section{New Physics}
\label{sec:kronfeld:new}

The foregoing discussion makes clear that it is desirable both for the 
experiments to improve further in precision and for the lattice-QCD 
calculations to be confirmed.
Given the current status, it is conceivable that the tension will 
increase again to the point that it warrants broad attention.
With that in mind, this section provides some information on 
extensions of the Standard Model.

The decays $D_s\to\ell\nu$ could be mediated by particles other 
than the Standard~$W$, either 
by $s$-channel annihilation via another charge-$+1$ particle, 
by $t$-channel exchange of a charge-$+\frac{2}{3}$ particle, or
by $u$-channel exchange of a charge-$-\frac{1}{3}$ particle.
All three kinds of particle are popular enough in extensions of the 
Standard Model to have their own sections in the Review of Particle 
Physics~\cite{Amsler:2008zzb,Yao:2006px}.
The charge-$+1$ particle would be a $W'$ or a charged Higgs boson;
the fractionally charged particles are known as leptoquarks.
All would have a mass, presumably, at least as large as~$M_W$.
Their interactions can be parametrized by the effective Lagrangian
\begin{eqnarray}
    \mathcal{L}_{\rm eff} & \hspace*{-6pt} = \hspace*{-6pt} & 
    \sqrt{2}G^\ell_A(\bar{s}\gamma^\mu\gamma_5c)(\bar{\nu}_L\gamma_\mu\ell_L) +
    \sqrt{2}G^\ell_P(\bar{s}\gamma_5c)(\bar{\nu}_L\ell_R) \nonumber \\ 
        & \hspace*{-6pt} - \hspace*{-6pt} &
    \sqrt{2}G^\ell_V(\bar{s}\gamma^\mu        c)(\bar{\nu}_L\gamma_\mu\ell_L) +
    \sqrt{2}G^\ell_S(\bar{s}        c)(\bar{\nu}_L\ell_R) \nonumber \\ 
        & \hspace*{-6pt} + \hspace*{-6pt} &
    \sqrt{2}G^\ell_T(\bar{s}\sigma^{\mu\nu}c)(\bar{\nu}_L\sigma_{\mu\nu}\ell_R),
    \label{eq:kronfeld:Leff}
\end{eqnarray}
where $G^l_A$ and $G^l_P$ appear in the leptonic-decay 
amplitude~(\ref{eq:kronfeld:ampBSM}).
The other interactions are likely to arise in non-Standard models, 
stemming from the chiral quantum numbers of the quarks and leptons.
Nonzero $G^\ell_V$ would interfere with the leading Standard amplitude 
of the semileptonic decay $D\to K\ell\nu$, potentially making a 
significant change in the rate~\cite{Kronfeld:2008gu}.
On the other hand, nonzero $G^\ell_S$ or $G^\ell_T$ would interfere 
with helicity-suppression, being visible only in an asymmetry of 
$D\to K\mu\nu$ after $10^7$ or more events are 
recorded~\cite{Kronfeld:2008gu}.

% Present update of bounds?

A $W'$ alters the (semi)leptonic amplitude via $G_A$ ($G_V$).
Barring a carefully-built (\emph{i.e.}, finely-tuned) model, this is 
not a promising scenario~\cite{Dobrescu:2008er}.
Many popular charged Higgs models are also unpromising.
Reference~\cite{Dobrescu:2008er} presents a charged Higgs model that 
could explain an excess of $D_s\to\ell\nu$ events, but it predicts the 
same-sized excess in $D^+\to\ell\nu$.
Now, however, this is disfavored by the near-perfect agreement the most 
precise measurement of $f_{D^+}$~\cite{:2008sq} with lattice 
QCD~\cite{Aubin:2005ar,Follana:2007uv,Bazavov:2009fDs,Blossier:2009bx}.

Leptoquarks, of several ilks, remain.
Even here the charge-$+\frac{2}{3}$ case is unpromising~\cite{Dobrescu:2008er}, 
owing to constraints from the lepton-flavor violating decays 
$\tau\to\mu\bar{s}s$, where $\bar{s}s$ hadronizes to $\phi$ or $KK$.
This leaves the most promising candidate to be an SU(2)-singlet, 
charged-$-\frac{1}{3}$ leptoquark.
This particle has the quantum numbers of a scalar down quark 
$\tilde{d}$, with an $R$-violating interaction
\begin{equation}
    (\kappa_{c\ell}\bar{c}_L\ell_L^c - 
        \kappa_{q\ell}V^*_{qs}\bar{s}_L \nu_{\ell L}^{\rm c})\tilde{d} +
    \kappa^\prime_{c\ell} \, \bar{c}_R\ell_R^{\rm c} \tilde{d} + 
    {\rm H.c.},
    \label{eq:kronfeld:leptoquark}
\end{equation}
where the superscript ``c'' denotes charge conjugation,
and $\kappa$ and $\kappa'$ are coupling matrices.
(With down squarks, the $\tilde{d}$ field should take a family index, 
and $\kappa$ and $\kappa'$ yet another index.)
Exchange of $\tilde{d}$ generates Eq.~(\ref{eq:kronfeld:Leff}) with
\begin{eqnarray}
    G^\ell_A = G^\ell_V & \hspace*{-6pt} = \hspace*{-6pt} &
        \kappa^*_{c\ell}\kappa_{ql}V^*_{qs}/4\sqrt{2}M_{\tilde{d}}^2, \\
    G^\ell_P = G^\ell_S & \hspace*{-6pt} = \hspace*{-6pt} &
        \kappa^{\prime*}_{c\ell}\kappa_{ql}V^*_{qs}/4\sqrt{2}M_{\tilde{d}}^2
    = 2G^\ell_T.
\end{eqnarray}
Generalizations of Eq.~(\ref{eq:kronfeld:leptoquark}) appear in 
non-Standard models that arise in many contexts~\cite{Kundu:2008ui,%
Dobrescu:2008sz,Benbrik:2008ik,Dey:2008ht,Hou:2008di,Fajfer:2008tm,%
Akeroyd:2009tn,Logan:2009uf,He:2009hz,Wei:2009nc,Mahmoudi:2009zx,%
Gninenko:2009yf,Deschamps:2009rh,Kao:2009mz,Tandean:2009yk,%
Bhattacharyya:2009hb}.

The leptoquark possibility been examined in a broader context 
including constraints from decays of the $D$ meson, kaon, 
$\tau$~lepton, and proton~\cite{Dorsner:2009cu}.
Reference~\cite{Dorsner:2009cu} claims these constraints make it 
difficult for the interaction~(\ref{eq:kronfeld:leptoquark}) to explain 
an enhancement in both $B(D_s\to\mu\nu)$ and $B(D_s\to\tau\nu)$ at the 
same time. 

\section{Summary}
\label{sec:kronfeld:summary}

\begin{figure}
    \includegraphics[width=\columnwidth]{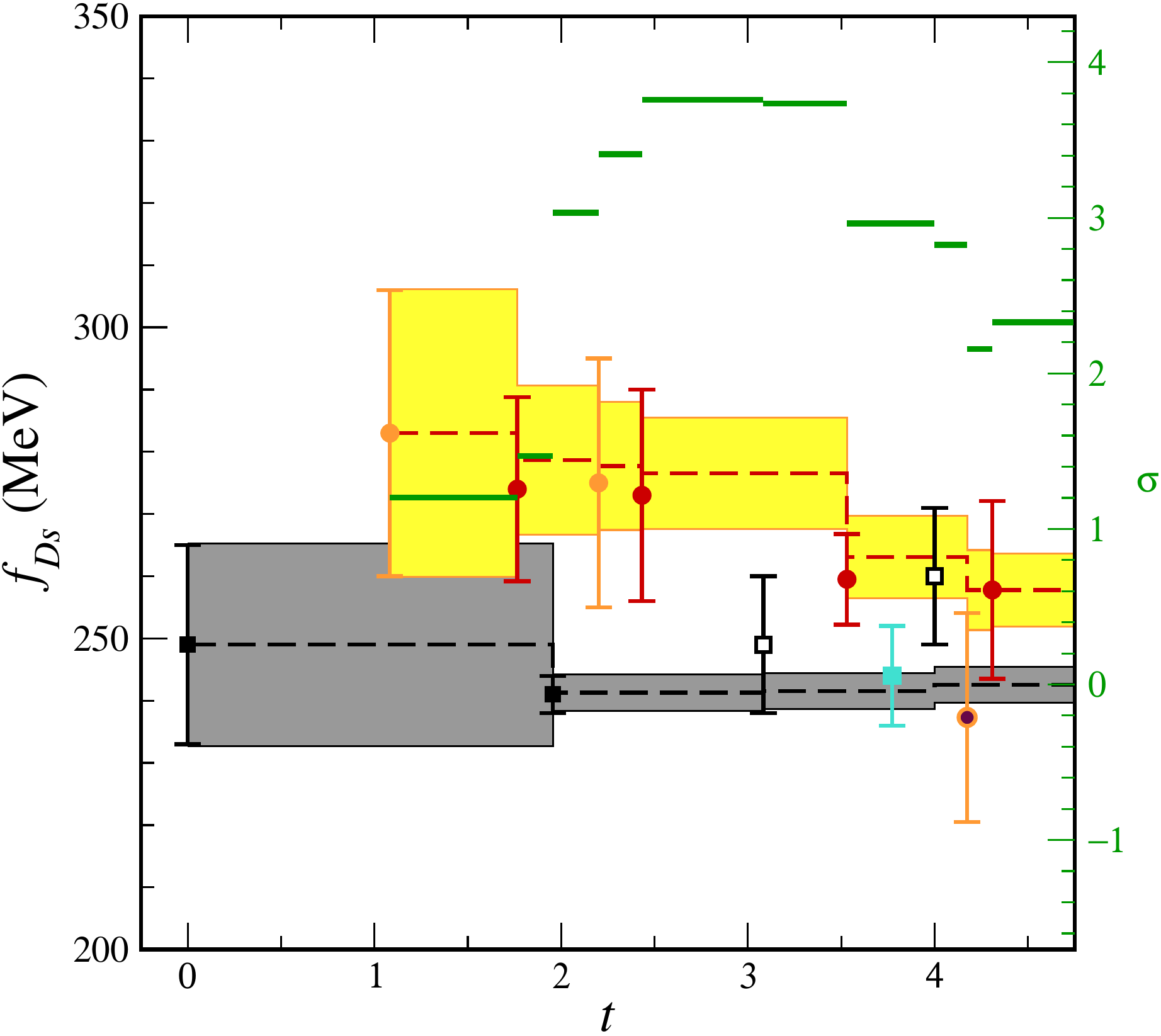}
    \caption[fig:kronfeld:2008]{
        Historical development of the $f_{D_s}$ puzzle since June 2005, 
        in years since Ref.~\cite{Aubin:2005ar} was posted at arXiv.org.
        Circles denote experimental measurements,
        from $e^+e^-$ collisions at the $\Upsilon(4S)$ in
        medium grey (orange);
        from $e^+e^-$ collisions at the $D_s^{(*)}D_s^{(*)}$ threshold 
        in dark grey (red);
        their weighted average is the medium grey (orange) dashed line
        with error band in light grey (yellow).
        Squares denote lattice-QCD calculations,
        with $n_f=2$ flavors in medium grey (cyan);
        with $n_f=2+1$ flavors in black (open symbols for conference reports);
        the $2+1$ weighted average is the black dashed line with error 
        band in grey.
        The discrepancy $\sigma$ is shown by the grey (green) solid 
        line segments, labeled by the vertical axis on the right.
        (color online)}
    \label{fig:kronfeld:2008}
%     \caption[fig:kronfeld:2008]{(color online)
%         Historical development of the $f_{D_s}$ puzzle since June 2005, 
%         in years since Ref.~\cite{Aubin:2005ar} was posted at arXiv.org.
%         Circles denote experimental measurements,
%         in $e^+e^-$ collisions at the $\Upsilon(4S)$ 
%         (orange or medium grey);
%         in $e^+e^-$ collisions at the $D_s^{(*)}D_s^{(*)}$ threshold 
%         (red or dark grey);
%         their weighted average is the orange (medium grey) dashed line
%         with error band in yellow (light grey).
%         Squares denote lattice-QCD calculations,
%         with $n_f=2$ flavors (cyan or medium grey);
%         with $n_f=2+1$ flavors (black; open symbols for conference 
%         reports);
%         the $2+1$ weighted average is the black dashed line with error 
%         band in grey.
%         The discrepancy $\sigma$ is shown by the  green (grey) solid 
%         line segments, labeled by the vertical axis on the right.}
\end{figure}

The developments of the $f_{D_s}$ puzzle are collected into
Fig.~\ref{fig:kronfeld:2008},
which presents the measurements and calculations discussed above, and 
the time dependence of their respective averages.
The time axis starts with the posting of Ref.~\cite{Aubin:2005ar}, the 
first lattice-QCD calculation with $2+1$ flavors of sea quarks.
The CLEO results of Ref.~\cite{Pedlar:2007za} ($t=1.8$) and of 
Refs.~\cite{Alexander:2009ux,Onyisi:2009th} ($t=3.5$) each are averaged 
to reduce clutter.

The discrepancy arose mostly with HPQCD's lattice-QCD calculation, 
and partly with the 2007 measurements from Belle and CLEO, rising to 
$3.8\sigma$.
In the past 18 months, the tension in $f_{D_s}$ has fallen to 
$2.3\sigma$, from
\begin{itemize}
    \item CLEO's new measurements of January 2009: $-0.8\sigma$;
    \item Fermilab/MILC's (preliminary) update for Lattice 2009: 
        $-0.13\sigma$;
    \item HFAG's reinterpretation of BaBar's measurement: 
        $-0.67\sigma$.
    \item CLEO's new measurement of October 2009: $+0.1\sigma$;
\end{itemize}
The history of the discrepancy's significance is traced via the grey 
(green) piece-wise horizontal line and right vertical axis
in Fig.~\ref{fig:kronfeld:2008}~~
Now, by the way, if one follows Ref.~\cite{Amsler:2008zzb} and omits 
the BaBar result (as reinterpreted by HFAG), 
the discrepancy is 2.6$\sigma$.

The prospects for a resolution of the $f_{D_s}$ puzzle are 
good---whether the tension goes away completely or tightens again.
BaBar is measuring the absolute branching ratio,
and Belle plans to update its analysis with higher statistics.
In a few years, BES~3 will measure $D_s\to\ell\nu$ in threshold 
production, similarly to CLEO, with a target uncertainty 
of 1\%~\cite{Li:2008wv}.
Several lattice-QCD collaborations now have enough $n_f=2+1$ ensembles
to carry out a useful calculation of $f_{D_s}$.
The MILC Collaboration has started to generate ensembles with 
$n_f=2+1+1$ sea quarks with the HISQ action, where the fourth sea quark 
is charm.
Similarly, the ETMC has embarked on a project with $n_f=2+1+1$ 
twisted-mass Wilson sea quarks.
Even if the puzzle dissipates, $D$ and $D_s$ leptonic decays will be 
useful for constraining extensions of the Standard Model~\cite{Dorsner:2009cu}.

\section*{Acknowledgments}

I would like to thank
Christine Davies,
Bogdan Dobrescu,
Alan Schwartz,
James Simone,
Sheldon Stone,
and
Ruth Van~de~Water
for fruitful discussions on the $f_{D_s}$ puzzle.
Fermilab is operated by Fermi Research Alliance, LLC, under 
Contract DE-AC02-07CH11359 with the US Department of Energy.


\begin{thebibliography}{99}
%
\bibitem{Marciano:2004uf}
  W.~J.~Marciano,
  %``Precise determination of |V(us)| from lattice calculations of  pseudoscalar
  %decay constants,''
  Phys.\ Rev.\ Lett.\  {\bf 93} (2004) 231803
  [arXiv:hep-ph/0402299];
  %%CITATION = PRLTA,93,231803;%%
%\cite{Antonelli:2008jg}\bibitem{Antonelli:2008jg}
  M.~Antonelli {\it et al.}  [FlaviaNet Working Group on Kaon Decays],
  %``Precision tests of the Standard Model with leptonic and semileptonic kaon
  %decays,''
  arXiv:0801.1817 [hep-ph].
  %%CITATION = ARXIV:0801.1817;%%
%
\bibitem{Briere:2001rn}
  R.~A.~Briere {\it et al.},
  ``{CLEO}-$c$ and {CESR}-$c$: A New Frontier of Weak and Strong 
  Interactions,''
  CLNS-01-1742.
  %%CITATION = CLNS-01-1742;%%
%
\bibitem{Davies:2003ik}
  C.~T.~H.~Davies {\it et al.}  [HPQCD, MILC, and Fermilab Lattice 
  Collaborations],
  %``High-precision lattice QCD confronts experiment,''
  Phys.\ Rev.\ Lett.\  {\bf 92} (2004) 022001
  [arXiv:hep-lat/0304004].
  %%CITATION = PRLTA,92,022001;%%
% 249 ± 3 ± 16 MeV (28 June 2005)
\bibitem{Aubin:2005ar}
  C.~Aubin {\it et al.} [Fermilab Lattice and MILC Collaborations],
  %``Charmed meson decay constants in three-flavor lattice QCD,''
  Phys.\ Rev.\ Lett.\  {\bf 95} (2005) 122002
  [arXiv:hep-lat/0506030].
  %%CITATION = PRLTA,95,122002;%%
% 283 ± 17 ± 7 ± 14 MeV (28 July 2006)
\bibitem{Aubert:2006sd}
  B.~Aubert {\it et al.} [BaBar Collaboration],
  %``Measurement of the pseudoscalar decay constant f(D/s) using 
  % charm-tagged events in e+ e- collisions at s**(1/2) = 10.58-GeV,''
  Phys.\ Rev.\ Lett.\  {\bf 98} (2007) 141801
  [arXiv:hep-ex/0607094].
  %%CITATION = PRLTA,98,141801;%%
% 264 ± 15 ± 7 (3 April 2007)
\bibitem{Pedlar:2007za}
  T.~K.~Pedlar {\it et al.}  [CLEO Collaboration],
  %``Measurement of B(D/s+ --> l+ nu) and the decay constant f(D/s+),''
  Phys.\ Rev.\  D {\bf 76} (2007) 072002
  [arXiv:0704.0437 [hep-ex]];
%%CITATION = PHRVA,D76,072002;%%
% 310 ± 25 ± 8 MeV \cite{Artuso:2007zg}\bibitem{Artuso:2007zg}
  M.~Artuso {\it et al.}  [CLEO Collaboration],
  %``Measurement of the decay constant $f_{D_s^+} using $D_s^+\to 
  %l^+\nu$,''
  Phys.\ Rev.\ Lett.\  {\bf 99} (2007) 071802
  [arXiv:0704.0629 [hep-ex]].
%%CITATION = PRLTA,99,071802;%%
% 275 ± 16 ±Ê12 MeV (10 September 2007)
\bibitem{Widhalm:2007mi}
  L.~Widhalm {\it et al.} [Belle Collaboration],
  %``Measurement of $B(D_s\to\mu\nu)$,''
  Phys.\ Rev.\ Lett.\  {\bf 100} (2008) 241801
  [arXiv:0709.1340 [hep-ex]].
  %%CITATION = PRLTA,100,241801;%%
% 273 ± 16 ±Ê8 MeV (7 December 2007)
\bibitem{Ecklund:2007zm}
  K.~M. Ecklund {\it et al.} [CLEO Collaboration],
  %``Measurement of the Absolute Branching Fraction of 
  %$D_s^+\arrow\tau^+\nu_\tau$ Decay,''
  Phys.\ Rev.\ Lett.\  {\bf 100} (2008) 161801
  [arXiv:0712.1175 [hep-ex]].
  %%CITATION = PRLTA,100,161801;%%
%
\bibitem{Dobrescu:2008er}
  B.~A.~Dobrescu and A.~S.~Kronfeld,
  %``Accumulating evidence for nonstandard leptonic decays of $D_s$ mesons,''
  Phys.\ Rev.\ Lett.\  {\bf 100} (2008) 241802
  [arXiv:0803.0512 [hep-ph]].
  %%CITATION = PRLTA,100,241802;%%
% 241 ± 3 MeV (12 June 2007)
\bibitem{Follana:2007uv}
  E.~Follana, C.~T.~H.~Davies, G.~P.~Lepage and J.~Shigemitsu 
  [HPQCD Collaboration],
  %``High precision determination of the pi, K, D and D_s decay 
  %constants from lattice QCD,''
  Phys. Rev. Lett. {\bf 100} (2008) 062002
  [arXiv:0706.1726 [hep-lat]].
%%CITATION = ARXIV:0706.1726;%%
%
\bibitem{Amsler:2008zzb}
  C.~Amsler {\it et al.}  [Particle Data Group],
  %``Review of particle physics,''
  Phys.\ Lett.\  B {\bf 667} (2008) 1
  %%CITATION = PHLTA,B667,1;%%
  [arXiv:0802.1043 [hep-ex]].
  %%CITATION = ARXIV:0802.1043;%%
% 232 ±45 ±20 ±48 MeV (3 August 1992)
\bibitem{Aoki:1992wj}
  S.~Aoki {\it et al.}  [WA75 Collaboration],
  %``The first observation of the muonic decay 
  %$D_s^\pm\to\mu^\pm\nu_\mu$,''
  Prog.\ Theor.\ Phys.\  {\bf 89} (1993) 131.
  %%CITATION = PTPKA,89,131;%%
% 344±37±52±42 MeV (3 August 1993)
\bibitem{Acosta:1993rx}
  D.~Acosta {\it et al.}  [CLEO Collaboration],
  %``First Measurement Of Gamma (D(S)+ $\to$ Mu+ Neutrino) / Gamma (D(S)+ $\to$
  %Phi Pi+),''
  Phys.\ Rev.\  D {\bf 49} (1994) 5690.
  %%CITATION = PHRVA,D49,5690;%%
% 430^{+150}_{-130}\pm 40  (22 December 1994)
\bibitem{Bai:1994qz}
  J.~Z.~Bai {\it et al.}  [BES Collaboration],
  %``A Direct measurement of the pseudoscalar decay constant, f(D(s)),''
  Phys.\ Rev.\ Lett.\  {\bf 74} (1995) 4599.
  %%CITATION = PRLTA,74,4599;%%
% 194 ± 35 ± 20 ± 14 MeV (30 May/16 June 1996)
\bibitem{Kodama:1996xq}
  K.~Kodama {\it et al.}  [Fermilab E653 Collaboration],
  %``Measurement of B(D_s+ -> mu+ nu_mu)/B(D_s+ -> phi mu+ nu_mu) and
  %Determination of the Decay Constant f_{D_s},''
  Phys.\ Lett.\  B {\bf 382} (1996) 299
  [arXiv:hep-ex/9606017].
  %%CITATION = PHLTA,B382,299;%%
% 309 ± 58 ± 33 ± 38 MeV (19 December 1996)
\bibitem{Acciarri:1996bv}
  M.~Acciarri {\it et al.}  [L3 Collaboration],
  %``Measurement of D/s- --> tau- anti-nu/tau and a new limit for B-  --> tau-
  %anti-nu/tau,''
  Phys.\ Lett.\  B {\bf 396} (1997) 327.
  %%CITATION = PHLTA,B396,327;%%
% 280 ± 17 ± 25 ± 34 MeV (11 December 1997/10 March 1998)
\bibitem{Chadha:1997zh}
  M.~Chadha {\it et al.}  [CLEO Collaboration],
  %``Improved measurement of the pseudoscalar decay constant f(D/s),''
  Phys.\ Rev.\  D {\bf 58} (1998) 032002
  [arXiv:hep-ex/9712014].
  %%CITATION = PHRVA,D58,032002;%%
% 323±44±12±34 MeV (7 February 2000)
\bibitem{Alexandrov:2000ns}
  Yu.~Alexandrov {\it et al.} [BEATRICE Collaboration],
  %``Measurement of the D/s --> mu nu/mu branching fraction and of the D/s
  %decay constant,''
  % 18 events
  Phys.\ Lett.\  B {\bf 478} (2000) 31.
  %%CITATION = PHLTA,B478,31;%%
% 286±44±41 MeV (9 March 2001)
\bibitem{Abbiendi:2001nb}
  G.~Abbiendi {\it et al.}  [OPAL Collaboration],
  %``Measurement of the branching ratio for D/s- --> tau- anti-nu/tau 
  %decays,''
  Phys.\ Lett.\  B {\bf 516} (2001) 236
  [arXiv:hep-ex/0103012].
  %%CITATION = PHLTA,B516,236;%%
% 285±19±40 MeV (8 January 2002)
\bibitem{Heister:2002fp}
  A.~Heister {\it et al.}  [ALEPH Collaboration],
  %``Leptonic decays of the D/s meson,''
  Phys.\ Lett.\  B {\bf 528} (2002) 1
  [arXiv:hep-ex/0201024].
  %%CITATION = PHLTA,B528,1;%%
%
\bibitem{Yao:2006px}
  W.~M.~Yao {\it et al.}  [Particle Data Group],
  %``Review of particle physics,''
  J.\ Phys.\ G {\bf 33} (2006) 1.
  %%CITATION = JPHGB,G33,1;%%
% 257.3 ± 10.3 ± 3.9 MeV; 278.7 ± 17.1 ± 3.8 MeV (9 January 2009)
\bibitem{Alexander:2009ux}
  J.~P.~Alexander {\it et al.}  [CLEO Collaboration],
  %``Measurement of $B{D_s^+ \to \ell^+ \nu}$ and the Decay Constant $fD_s^+$
  %From 600 $/pb^{-1}$ of $e^\pm$ Annihilation Data Near 4170 MeV,''
  Phys.\ Rev.\  D {\bf 79} (2009) 052001
  [arXiv:0901.1216 [hep-ex]].
  %%CITATION = PHRVA,D79,052001;%%
% 252.5 ± 11.1 ± 5.2 MeV (9 January 2009)
\bibitem{Onyisi:2009th}
  P.~U.~E.~Onyisi {\it et al.}  [CLEO Collaboration],
  %``Improved Measurement of Absolute Branching Fraction of Ds to tau nu,''
  Phys.\ Rev.\  D {\bf 79} (2009) 052002
  [arXiv:0901.1147 [hep-ex]].
  %%CITATION = PHRVA,D79,052002;%%
% 257.8 ± 13.3 ± 5.2 MeV (19 October 2009)
\bibitem{Naik:2009tk}
  P. Naik {\it et al.} [CLEO Collaboration],
  %``Measurement of the Pseudoscalar Decay Constant fDs Using Ds+ -> tau+ nu,
  %tau+ -> rho+ anti-nu Decays,''
  arXiv:0910.3602 [hep-ex].
  %%CITATION = ARXIV:0910.3602;%%
%
\bibitem{Aubert:2005xu}
  B.~Aubert {\it et al.}  [BaBar Collaboration],
  %``Measurement of the $B^0 \to D^{*-}D_s^{*+}$ and $D_s^+\to\phi\pi^+$
  %branching  fractions,''
  Phys.\ Rev.\  D {\bf 71} (2005) 091104
  [arXiv:hep-ex/0502041];
  %%CITATION = PHRVA,D71,091104;%%
%\cite{Aubert:2006nm}\bibitem{Aubert:2006nm}
%   B.~Aubert {\it et al.}  [BABAR Collaboration],
  %``Study of $B \to D^{(*)} D^{(*)}_{sJ}$ Decays and Measurement of
  %$D^-_{s}$ and $D_{sJ}(2460)$---Branching Fractions,''
  Phys.\ Rev.\  D {\bf 74} (2006) 031103
  [arXiv:hep-ex/0605036].
  %%CITATION = PHRVA,D74,031103;%%
%
\bibitem{Alexander:2008cq}
  J. Alexander {\it et al.}  [CLEO Collaboration],
  %``Absolute measurement of hadronic branching fractions of the 
  % $D_s^+$ meson,''
  Phys.\ Rev.\ Lett.\  {\bf 100} (2008) 161804
  [arXiv:0801.0680 [hep-ex]].
  %%CITATION = ARXIV:0801.0680;%%
%
\bibitem{Burdman:1994ip}
  G.~Burdman, J.~T.~Goldman and D.~Wyler,
  %``Radiative leptonic decays of heavy mesons,''
  Phys.\ Rev.\  D {\bf 51} (1995) 111
  [arXiv:hep-ph/9405425].
  %%CITATION = PHRVA,D51,111;%%
%
\bibitem{Hwang:2005uk}
  C.~W.~Hwang,
  %``Radiative leptonic decays of heavy mesons in heavy quark limit,''
  Eur.\ Phys.\ J.\  C {\bf 46} (2006) 379
  [arXiv:hep-ph/0512006].
  %%CITATION = EPHJA,C46,379;%%
%
\bibitem{Schwartz:2009hv}
  A.~J.~Schwartz,
  %``B+ and Ds+ Decay Constants from Belle and Babar,''
  proceedings of the Xth Conference on the Intersections of
  Particle and Nuclear Physics (AIP, Melville~NY, 2009)
  [arXiv:0909.4473 [hep-ex]].
  %%CITATION = ARXIV:0909.4473;%%
%
\bibitem{HFAG}
  Heavy Flavor Averaging Group (HFAG), 
  http://www.slac.stanford.edu/xorg/hfag/charm/.
%
\bibitem{Peters:2009jy}
  K.~Peters,
  %``Higgs Searches,''
  these proceedings,
  arXiv:0911.1469 [hep-ex].
  %%CITATION = ARXIV:0911.1469;%%
%
\bibitem{Bazavov:2009bb}
  A.~Bazavov {\it et al.},
  %``Full nonperturbative QCD simulations with 2+1 flavors of improved 
  %staggered quarks,''
  arXiv:0903.3598 [hep-lat];
  %%CITATION = ARXIV:0903.3598;%%
  S.~Hashimoto, these proceedings.
%
\bibitem{Kronfeld:2002nm}
  A.~S.~Kronfeld,
  ``Progress in Lattice QCD,'' in the proceedings of the
  \emph{XX$^{\rm nd}$ Physics in Collision}, 
  (Stanford CA, 20--22 Jun 2002, eConf C020620)
  [arXiv:hep-ph/0209231].
  %%CITATION = ECONF,C020620,FRBT05;%%
%
\bibitem{Bernard:2001av}
  C.~W.~Bernard {\it et al.} [MILC Collaboration],
  %``The QCD spectrum with three quark flavors,''
  Phys.\ Rev.\  D {\bf 64}, 054506 (2001)
  [arXiv:hep-lat/0104002];
  %%CITATION = PHRVA,D64,054506;%%
%\cite{Aubin:2004wf}\bibitem{Aubin:2004wf}
  C.~Aubin {\it et al.} [MILC Collaboration],
  %``Light hadrons with improved staggered quarks: Approaching the continuum
  %limit,''
  Phys.\ Rev.\  D {\bf 70}, 094505 (2004)
  [arXiv:hep-lat/0402030].
  %%CITATION = PHRVA,D70,094505;%%
%
\bibitem{Kronfeld:2007ek}
  See, for example, A.~S.~Kronfeld,
  %``Lattice gauge theory with staggered fermions: how, where, and why (not),''
  PoS {\bf LATTICE 2007} (2007) 016
  [arXiv:0711.0699 [hep-lat]] and references therein.
  %%CITATION = POSCI,LAT2007,016;%%
%
\bibitem{Shamir:2004zc}
  Y.~Shamir,
  %``Locality of the fourth root of the staggered-fermion determinant:
  %Renormalization-group approach,''
  Phys.\ Rev.\  D {\bf 71} (2005) 034509
  [arXiv:hep-lat/0412014];
  %%CITATION = PHRVA,D71,034509;%%
%\cite{Bernard:2006ee}\bibitem{Bernard:2006ee}
  C.~Bernard, M.~Golterman and Y.~Shamir,
  %``Observations on staggered fermions at non-zero lattice spacing,''
  Phys.\ Rev.\  D {\bf 73} (2006) 114511
  [arXiv:hep-lat/0604017];
  %%CITATION = PHRVA,D73,114511;%%
%\cite{Shamir:2006nj}\bibitem{Shamir:2006nj}  
  Y.~Shamir,
  %``Renormalization-group analysis of the validity of staggered-fermion QCD
  %with the fourth-root recipe,''
  Phys.\ Rev.\  D {\bf 75} (2007) 054503
  [arXiv:hep-lat/0607007].
  %%CITATION = PHRVA,D75,054503;%%
%
\bibitem{Nielsen:1980rz}
  H.~B.~Nielsen and M.~Ninomiya,
  %``Absence Of Neutrinos On A Lattice. 1. Proof By Homotopy Theory,''
  Nucl.\ Phys.\  B {\bf 185} (1981) 20; (E) {\bf 195} (1982) 541.
  %%CITATION = NUPHA,B185,20;%%
%
\bibitem{Susskind:1976jm}
  L.~Susskind,
  %``Lattice Fermions,''
  Phys.\ Rev.\  D {\bf 16} (1977) 3031.
  %%CITATION = PHRVA,D16,3031;%%
%
\bibitem{Hamber:1983kx}
  H.~W.~Hamber, E.~Marinari, G.~Parisi and C.~Rebbi,
  %``Numerical Simulations Of Quantum Chromodynamics,''
  Phys.\ Lett.\  B {\bf 124} (1983) 99.
  %%CITATION = PHLTA,B124,99;%%
%
\bibitem{Aubin:2003mg}
  C.~Aubin and C.~Bernard,
  %``Pion and Kaon masses in Staggered Chiral Perturbation Theory,''
  Phys.\ Rev.\  D {\bf 68} (2003) 034014
  [arXiv:hep-lat/0304014];
  %%CITATION = PHRVA,D68,034014;%%
%\cite{Aubin:2003uc}\bibitem{Aubin:2003uc}  C.~Aubin and C.~Bernard,
  %``Pseudoscalar decay constants in staggered chiral perturbation theory,''
  Phys.\ Rev.\  D {\bf 68} (2003) 074011
  [arXiv:hep-lat/0306026].
  %%CITATION = PHRVA,D68,074011;%%
%
\bibitem{Bernard:2007ps}
  C.~Bernard {\it et al.},
  %``Status of the MILC light pseudoscalar meson project,''
  PoS {\bf LATTICE 2007} (2007) 090
  [arXiv:0710.1118 [hep-lat]].
  %%CITATION = POSCI,LAT2007,090;%%
%
\bibitem{Follana:2006rc}
  E.~Follana {\it et al.}  [HPQCD Collaboration and UKQCD Collaboration],
  %``Highly improved staggered quarks on the lattice, with applications
  %to charm physics,''
  Phys.\ Rev.\  D {\bf 75} (2007) 054502
  [arXiv:hep-lat/0610092].
  %%CITATION = PHRVA,D75,054502;%%
%
\bibitem{Smit:1987zh}
  J.~Smit and J.~C.~Vink,
  %``Renormalized Ward-Takahashi relations and topological susceptibility with
  %staggered fermions,''
  Nucl.\ Phys.\  B {\bf 298} (1988) 557.
  %%CITATION = NUPHA,B298,557;%%
%
\bibitem{ElKhadra:1996mp}
  A.~X.~El-Khadra, A.~S.~Kronfeld and P.~B.~Mackenzie,
  %``Massive Fermions in Lattice Gauge Theory,''
  Phys.\ Rev.\  D {\bf 55} (1997) 3933
  [arXiv:hep-lat/9604004];
  %%CITATION = PHRVA,D55,3933;%%
%\cite{Kronfeld:2000ck}\bibitem{Kronfeld:2000ck}
  A.~S.~Kronfeld,
  %``Application of heavy-quark effective theory to lattice QCD. I: Power
  %corrections,''
  Phys.\ Rev.\  D {\bf 62} (2000) 014505
  [arXiv:hep-lat/0002008].
  %%CITATION = PHRVA,D62,014505;%%
%
\bibitem{Harada:2001fi}
  J.~Harada, S.~Hashimoto, K.~I.~Ishikawa, A.~S.~Kron\-feld, T.~Onogi and 
  N.~Yamada,
  %``Application of heavy-quark effective theory to lattice QCD. II:  
  %Radiative corrections to heavy-light currents,''
  Phys.\ Rev.\  D {\bf 65} (2002) 094513 [arXiv:hep-lat/0112044];
  (E) {\bf 71} (2005) 019903;
  %%CITATION = PHRVA,D65,094513;%%
%\cite{ElKhadra:2007qe}\bibitem{ElKhadra:2007qe}
  A.~X.~El-Khadra, E.~G\'amiz, A.~S.~Kronfeld and M.~A.~Nobes,
  %``Perturbative matching of heavy-light currents at one-loop,''
  PoS {\bf LATTICE 2007} (2007) 242
  [arXiv:0710.1437 [hep-lat]].
  %%CITATION = POSCI,LAT2007,242;%%
%
\bibitem{Kronfeld:2003sd}
  A.~S.~Kronfeld,
  %``Heavy quarks and lattice QCD,''
  Nucl.\ Phys.\ Proc.\ Suppl.\  {\bf 129} (2004) 46
  [arXiv:hep-lat/0310063].
  %%CITATION = NUPHZ,129,46;%%
%
\bibitem{:2008sq}
  B.~I.~Eisenstein {\it et al.}  [CLEO Collaboration],
  %``Precision Measurement of B(D+ -> mu+ nu) and the Pseudoscalar Decay
  %Constant fD+,''
  Phys.\ Rev.\  D {\bf 78} (2008) 052003
  [arXiv:0806.2112 [hep-ex]].
  %%CITATION = PHRVA,D78,052003;%%
%
\bibitem{Sommer:1993ce}
  R.~Sommer,
  %``A new way to set the energy scale in lattice gauge theories and its
  %applications to the static force and alpha-s in SU(2) Yang-Mills theory,''
  Nucl.\ Phys.\  B {\bf 411} (1994) 839
  [arXiv:hep-lat/9310022].
  %%CITATION = NUPHA,B411,839;%%
%
\bibitem{Bernard:2000gd}
  C.~W.~Bernard {\it et al.},
  %``The static quark potential in three flavor QCD,''
  Phys.\ Rev.\  D {\bf 62} (2000) 034503
  [arXiv:hep-lat/0002028].
  %%CITATION = PHRVA,D62,034503;%%
%
\bibitem{Gray:2005ur}
  A.~Gray \emph{et al.} [HPQCD Collaboration], 
  % I.~Allison, C.~T.~H.~Davies, E.~Dalgic, G.~P.~Lepage, 
  % J.~Shigemitsu and M.~Wingate,
  %``The $\Upsilon$ spectrum and $m_b$ from full lattice QCD,''
  Phys.\ Rev.\  D {\bf 72} (2005) 094507
  [arXiv:hep-lat/0507013].
  %%CITATION = PHRVA,D72,094507;%%
% 249 ± 11 MeV July 2008
\bibitem{Bernard:2009wr}
  C.~Bernard {\it et al.} [Fermilab Lattice and MILC Collaborations],
  %``B and D Meson Decay Constants,''
  PoS {\bf LATTICE 2008} (2008) 278
  [arXiv:0904.1895 [hep-lat]].
  %%CITATION = ARXIV:0904.1895;%%
% 260 ± 10 MeV (Lattice 2009)
\bibitem{Bazavov:2009fDs}
  A.~Bazavov {\it et al.}  [Fermilab Lattice and MILC Collaborations],
  %``The $D_s$ and $D^+$ Leptonic Decay Constants from Lattice QCD,''
  PoS {\bf LATTICE 2009} (2009) 249.
  %%CITATION = POSCI,LATTICE2009,249;%%
%
\bibitem{Bazavov:2009fk}
  A.~Bazavov {\it et al.}  [MILC Collaboration],
  %``MILC results for light pseudoscalars,''
  PoS {\bf CD09} (2009) 007
  [arXiv:0910.2966 [hep-ph]];
  %%CITATION = ARXIV:0910.2966;%%
  PoS {\bf LATTICE 2009} (2009) 079
  [arXiv:0910.3618 [hep-lat]].
  %%CITATION = ARXIV:0910.3618;%%
%
\bibitem{Davies:2009ts}
  C.~T.~H.~Davies, E.~Follana, I.~D.~Kendall, G.~P. Lepage and C.~McNeile,
  %``Precise determination of the lattice spacing in full lattice QCD,''
  arXiv:0910.1229 [hep-lat].
  %%CITATION = ARXIV:0910.1229;%%
% 244 ± 8 MeV (nf = 2, 6 April 2009)
\bibitem{Blossier:2009bx}
  B.~Blossier {\it et al.} [European Twisted-Mass Collaboration],
  %``Pseudoscalar decay constants of kaon and $D$ mesons from $N_f=2$ 
  %twisted-mass lattice QCD,''
  JHEP {\bf 0907} (2009) 043
  [arXiv:0904.0954 [hep-lat]].
  %%CITATION = JHEPA,0907,043;%%
% 241 ± 5 ^{+27+9+5}_{26-4-0} MeV (17 June 2002)
\bibitem{Bernard:2002pc}
  C.~Bernard {\it et al.}  [MILC Collaboration],
  %``Lattice calculation of heavy-light decay constants with two flavors of
  %dynamical quarks,''
  Phys.\ Rev.\  D {\bf 66} (2002) 094501
  [arXiv:hep-lat/0206016].
  %%CITATION = PHRVA,D66,094501;%%
%267(13)(17)(^{+10}_{-0}) MeV for Nf=2
\bibitem{AliKhan:2000eg}
  A.~Ali Khan {\it et al.}  [CP-PACS Collaboration],
  %``Decay constants of B and D mesons from improved relativistic lattice  QCD
  %with two flavours of sea quarks,''
  Phys.\ Rev.\  D {\bf 64} (2001) 034505
  [arXiv:hep-lat/0010009].
  %%CITATION = PHRVA,D64,034505;%%
%
\bibitem{Kronfeld:2008gu}
  A.~S.~Kronfeld,
  %``Non-Standard Physics in Leptonic and Semileptonic Decays of Charmed
  %Mesons,''
  PoS {\bf LATTICE 2008} (2008) 282
  [arXiv:0812.2030 [hep-lat]].
  %%CITATION = ARXIV:0812.2030;%%
%
\bibitem{Kundu:2008ui}
  A.~Kundu and S.~Nandi,
  %``R-parity violating supersymmetry, $B_s$ mixing, and $D_s \to \ell \nu$,''
  Phys.\ Rev.\  D {\bf 78} (2008) 015009
  [arXiv:0803.1898 [hep-ph]].
  %%CITATION = PHRVA,D78,015009;%%
%
\bibitem{Dobrescu:2008sz}
  B.~A.~Dobrescu and P.~J.~Fox,
  %``Quark and lepton masses from top loops,''
  JHEP {\bf 0808} (2008) 100
  [arXiv:0805.0822 [hep-ph]].
  %%CITATION = JHEPA,0808,100;%%
%
\bibitem{Benbrik:2008ik}
  R.~Benbrik and C.~H.~Chen,
  %``Leptoquark on $P\to \ell^{+} \nu$, FCNC and LFV,''
  Phys.\ Lett.\  B {\bf 672} (2009) 172
  [arXiv:0807.2373 [hep-ph]].
  %%CITATION = PHLTA,B672,172;%%
%
\bibitem{Dey:2008ht}
  P.~Dey, A.~Kundu, B.~Mukhopadhyaya and S.~Nandi,
  %``Two-loop neutrino masses with large R-parity violating interactions in
  %supersymmetry,''
  JHEP {\bf 0812} (2008) 100
  [arXiv:0808.1523 [hep-ph]].
  %%CITATION = JHEPA,0812,100;%%
%
\bibitem{Hou:2008di}
  G.~W.~S.~Hou,
  %``Search for TeV Scale Physics in Heavy Flavour Decays,''
  Eur.\ Phys.\ J.\  C {\bf 59} (2009) 521
  [arXiv:0808.1932 [hep-ex]].
  %%CITATION = EPHJA,C59,521;%%
%
\bibitem{Fajfer:2008tm}
  S.~Fajfer and N.~Ko\v snik,
  %``Leptoquarks in FCNC charm decays,''
  Phys.\ Rev.\  D {\bf 79} (2009) 017502
  [arXiv:0810.4858 [hep-ph]].
  %%CITATION = PHRVA,D79,017502;%%
%
\bibitem{Akeroyd:2009tn}
  A.~G.~Akeroyd and F.~Mahmoudi,
  %``Constraints on charged Higgs bosons from D(s)+- -> mu+- nu and D(s)+- ->
  %tau+- nu,''
  JHEP {\bf 0904} (2009) 121
  [arXiv:0902.2393 [hep-ph]].
  %%CITATION = JHEPA,0904,121;%%
%
\bibitem{Logan:2009uf}
  H.~E.~Logan and D.~MacLennan,
  %``Charged Higgs phenomenology in the lepton-specific two Higgs doublet
  %model,''
  Phys.\ Rev.\  D {\bf 79} (2009) 115022
  [arXiv:0903.2246 [hep-ph]].
  %%CITATION = PHRVA,D79,115022;%%
%
\bibitem{He:2009hz}
  X.~G.~He, J.~Tandean and G.~Valencia,
  %``Probing New Physics in Charm Couplings with FCNC,''
  Phys.\ Rev.\  D {\bf 80} (2009) 035021
  [arXiv:0904.2301 [hep-ph]].
  %%CITATION = PHRVA,D80,035021;%%
%
\bibitem{Wei:2009nc}
  Z.~T.~Wei, H.~W.~Ke and X.~F.~Yang,
  %``Interpretation of the '$f_{D_s}$ puzzle' in SM and beyond,''
  Phys.\ Rev.\  D {\bf 80} (2009) 015022
  [arXiv:0905.3069 [hep-ph]].
  %%CITATION = PHRVA,D80,015022;%%
%
\bibitem{Mahmoudi:2009zx}
  F. Mahmoudi and O. St\aa l,
  %``Flavor constraints on the two-Higgs-doublet model with general Yukawa
  %couplings,''
  arXiv:0907.1791 $\!$[hep-ph].
  %%CITATION = ARXIV:0907.1791;%%
%
\bibitem{Gninenko:2009yf}
  S.~N.~Gninenko and D.~S.~Gorbunov,
  %``The MiniBooNE anomaly, the decay Ds -> mu+nu and heavy sterile neutrino,''
  arXiv:0907.4666 [hep-ph].
  %%CITATION = ARXIV:0907.4666;%%
%
\bibitem{Deschamps:2009rh}
  O.~Deschamps, S.~Descotes-Genon, S.~Monteil, V.~Niess, S.~T'Jampens 
  and V.~Tisserand,
  %``The Two Higgs Doublet of Type II facing flavour physics data,''
  arXiv:0907.5135 [hep-ph].
  %%CITATION = ARXIV:0907.5135;%%
%
\bibitem{Kao:2009mz}
  Y.~Kao and T.~Takeuchi,
  %``Constraints on R-parity violation from recent Belle/Babar data,''
  arXiv:0909.0042 [hep-ph].
  %%CITATION = ARXIV:0909.0042;%%
%
\bibitem{Tandean:2009yk}
  J.~Tandean,
  %``Probing New Physics in Charm Couplings with Kaon and Other Hadron
  %Processes,''
  arXiv:0909.3957 [hep-ph].
  %%CITATION = ARXIV:0909.3957;%%
%
\bibitem{Bhattacharyya:2009hb}
  G.~Bhattacharyya, K.~B.~Chatterjee and S.~Nandi,
  %``Correlated enhancements in $D_s \to \ell\nu$, $(g-2)$ of muon, and lepton
  %flavor violating $\tau$ decays with two R-parity violating couplings,''
  arXiv:0911.3811 [hep-ph].
  %%CITATION = ARXIV:0911.3811;%%
%
\bibitem{Dorsner:2009cu}
  I.~Dor\v sner, S.~Fajfer, J.~F.~Kamenik and N.~Ko\v snik,
  %``Can scalar leptoquarks explain the f_{D_s} puzzle?,''
  Phys.\ Lett.\  B {\bf 682} (2009) 67
  [arXiv:0906.5585 [hep-ph]].
  %%CITATION = PHLTA,B682,67;%%
%
\bibitem{Li:2008wv}
  H.~B.~Li and J.~H.~Zou,
  %``A possible signature of new physics at BES-III,''
  Chin.\ Phys.\  C {\bf 33} (2009)~1
  [arXiv:0804.1822 [hep-ex]];
  %%CITATION = CHPHD,C33,1;%%
%\cite{Asner:2008nq}\bibitem{Asner:2008nq}
  D.~M.~Asner {\it et al.},
  ``Physics at BES~III,''
  arXiv:0809.1869 [hep-ex].
  %%CITATION = ARXIV:0809.1869;%%
\end{thebibliography}
\end{document}